\documentclass{elsart}
\usepackage{amsfonts}
\usepackage{amsmath}
\usepackage{epsfig}
\usepackage{amssymb}
\usepackage[figuresright]{rotating}

\newcommand{\fpi}{f_\pi}
\newcommand{\mpi}{m_\pi}
\newcommand{\mev}{\,{\rm MeV}}
\newcommand{\gev}{\,{\rm GeV}}
\newcommand{\fm}{\,{\rm fm}}

\newcommand{\be}{\begin{equation}}
\newcommand{\ee}{\end{equation}}
\newcommand{\ba}{\begin{array}}
\newcommand{\ea}{\end{array}}
\newcommand{\bea}{\begin{eqnarray}}
\newcommand{\eea}{\end{eqnarray}}

\newcommand{\err}[2]{${\scriptstyle {}^{+{#1}}_{-{#2}}}$}
\newcommand{\er}[2]{{\scriptstyle {}^{+{#1}}_{-{#2}}}}

\begin{document}


\begin{frontmatter}

\title{Chiral and Continuum Extrapolation of Partially-Quenched Lattice Results}

\author[swan]{C.~R.~Allton},
\author[swan]{W.~Armour},
\author[cssm]{D.~B.~Leinweber},
\author[jlab]{A.~W.~Thomas},
\author[jlab]{R.~D.~Young}
\address[swan]{Department of Physics, University of Wales Swansea, Swansea
SA2~8PP, Wales, U.K.}
\address[cssm]{CSSM and Department of Physics, University of Adelaide, Adelaide SA 5005, Australia}
\address[jlab]{Jefferson Lab, 12000 Jefferson Ave., Newport News VA 23606, USA}


\begin{abstract}
The vector meson mass is extracted from a large sample of partially
quenched, two-flavor lattice QCD simulations.  For the first time,
discretisation, finite-volume and partial quenching artefacts are
treated in a unified framework which is consistent with the low-energy
behaviour of QCD. This analysis incorporates the leading infrared
behaviour dictated by chiral effective field theory.  As the
two-pion decay channel cannot be described by a low-energy
expansion alone, a
highly-constrained model for the decay channel of the rho-meson is
introduced.  The latter
is essential for extrapolating lattice results from the quark-mass
regime where the rho is observed to be a physical bound state.

\end{abstract}

\end{frontmatter}

\begin{figure}[!t]
\begin{center}
\includegraphics[width=12cm]{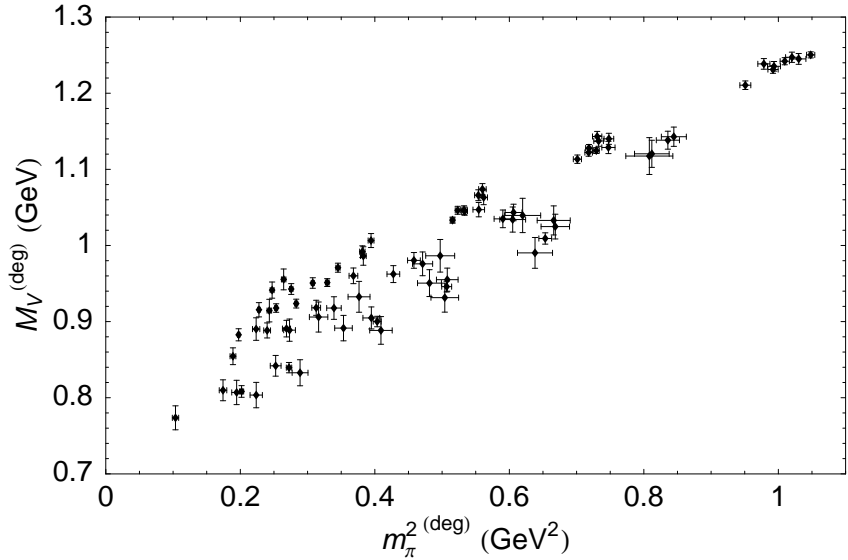}
\end{center}
\caption{Partially-quenched vector meson masses plotted versus
``degenerate'' pion mass squared. Here $M_V^{(deg)} = M^a_{ijj}$
and $m_\pi^{(deg)} = m^a_{ijj}$.  The simulation results are from CP-PACS
\cite{AliKhan:2001tx}.
\label{fig:data}}
\end{figure}
\begin{figure}[!t]
\begin{center}
\includegraphics[width=12cm]{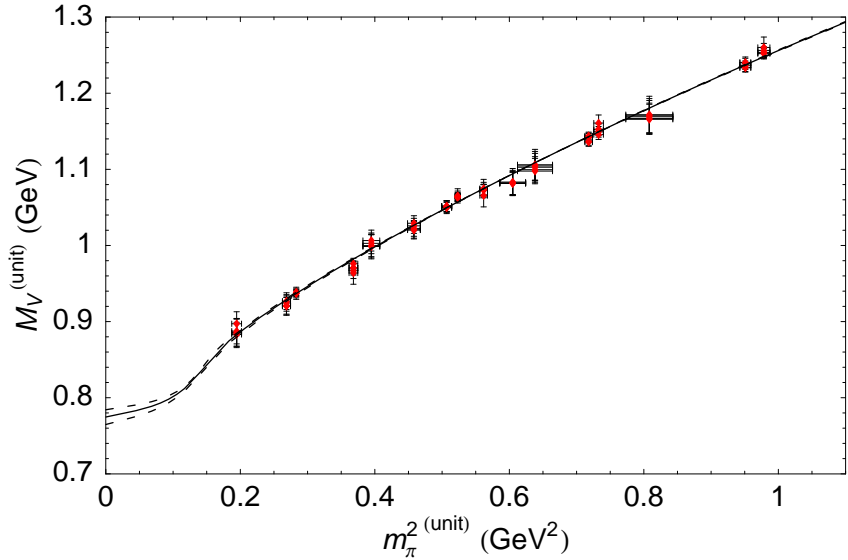}
\end{center}
\caption{The same 80 lattice data points as in Fig.~2, 
after correction to restore the infinite-volume,
continuum and quark-mass unitarity limits. The central curve displays the
best-fit from the global analysis. The dashed curves show the bounds
on the FRR scale, $0.620<\Lambda<0.690\gev$.
\label{fig:twist}}
\end{figure}

Recent developments in lattice QCD have enabled the first large-scale
simulation of chiral, dynamical 
fermions \cite{Aoki:2004ht}. While this
accomplishment is a significant milestone in the progress towards
an accurate description of physical QCD, the high demand on computing
resources restricts practical calculations to an unphysical domain of
simulation parameters. In particular, lattice QCD involves a
discretised space-time of finite spatial extent, with input quark
masses much larger than those in Nature. Each of these approximations
requires special attention in the extraction of physical observables
from Monte Carlo simulations.
In this Letter we analyse a very large set of partially quenched data for the 
mass of the $\rho$ meson. We show that 
a systematic analysis of this data 
enables us to remove the effects of  
partial quenching and to take both the 
continuum and infinite volume limits. The resulting data lies on a single, 
well defined curve which extrapolates to a value within $\sim$1\% of the physical 
$\rho$ mass. The contrast between the raw lattice data (note the scatter in 
Fig.~\ref{fig:data}) and the corrected data shown in Fig.~\ref{fig:twist}
is striking.

Spontaneous chiral symmetry breaking in QCD dictates that, in the
vicinity of the chiral limit, hadronic observables exhibit nonanalytic
dependence on the quark mass \cite{Li:1971vr}. This feature places
tight constraints on the form of chiral extrapolations if they are to
be consistent with the properties of low-energy QCD
\cite{Leinweber:1998ej,Leinweber:1999ig}.  The most natural solution
to this problem is to use effective field theory (EFT) to describe the
quark-mass dependence of hadron properties.  Considering a benchmark
quantity, such as the nucleon mass, there is substantial
phenomenological information on the quark-mass expansion near the
chiral limit \cite{Borasoy:1996bx}. In the context of lattice
simulations, where quark masses are significantly far from the chiral
limit, the expansion is acutely sensitive to higher-order
corrections. Fortunately, such issues can be alleviated by
reformulating the EFT in the framework of finite-range regularisation
(FRR) \cite{Young:2002ib} --- with demonstrated success in the
efficient extrapolation \cite{Leinweber:2003dg} of lattice
calculations of the nucleon mass \cite{AliKhan:2001tx}.

Provided simulations are performed on a suitably large box,
finite-volume corrections will be exponentially
suppressed. Nevertheless, these leading corrections can be described
by the same low-energy effective field theory used to understand the
quark-mass variation \cite{Gasser:1987zq}. Finite-volume corrections
are dominated by the suppression of the infrared component of chiral
loop diagrams --- as observed in a recent study of volume dependence
in lattice QCD \cite{AliKhan:2003cu} (using the quark-mass dependence
described in Ref.~\cite{Procura:2003ig}). Corrections of this type
have previously been incorporated in quark-mass extrapolations
\cite{Leinweber:2001ac,Young:2002cj,Young:2004tb}. They are essential
in the case of the $p$-wave decay channels, such as $\rho\to\pi\pi$
\cite{Leinweber:2001ac} and $\Delta\to N\pi$ \cite{Young:2002cj}.  The
modifications to EFT on a finite volume have been investigated for a
range of observables --- e.g., see
Refs.~\cite{Colangelo:2003hf,Beane:2004tw,Beane:2004rf,Leinweber:2004tc,%
Borasoy:2004zf}.

Removal of discretisation artefacts from simulation results also
represents an important step in the systematic extraction of continuum
QCD physics. From a technical point of view, a great deal of effort
has gone into action improvement \cite{Luscher:1996ug} such that
near-continuum results can be obtained at finite lattice spacing
\cite{Edwards:1997nh,Dong:2000mr,Zanotti:2004dr}. Residual
discrepancies from the continuum can be incorporated as perturbative
corrections in EFT
\cite{Rupak:2002sm,Bar:2003mh,Aoki:2003yv,Beane:2003xv}, thereby
providing a systematic approach to the continuum.

The generation of gauge field configurations with dynamical sea quarks
is the most computationally demanding component of the calculation of
standard observables. By comparison, the matrix inversion required to 
obtain quark propagators is relatively efficient. This enables the
calculation of quark propagators over a range of quark masses for a
fixed gauge field ensemble.  Such calculations are referred to as
partially quenched QCD (pQQCD), where the valence quark masses no
longer match those simulated in the sea. Although an unphysical
approximation, the connection to the physical theory in EFT has been
demonstrated \cite{Sharpe:2000bc}. Most importantly, the
partially quenched EFT does not require any new, unphysical parameters.

%
%
While EFT provides a systematic framework for the analysis of lattice
results, the present analysis of the $\rho$ meson requires one to go
beyond the scope of EFT. Near the chiral limit, the $\rho$ decays to
two energetic pions. The pions contributing to the imaginary part of
the $\rho$ mass cannot be considered soft, and therefore cannot be
systematically incorporated into a low-energy counting scheme
\cite{Bijnens:1997ni,Bruns:2004tj}. Because almost all the lattice
simulation points in this analysis lie in the region $\mpi>m_\rho/2$,
it is evident that in the extrapolation to the chiral regime will
encounter a threshold effect where the decay channel opens. To
incorporate this physical threshold, we model the $\rho\to\pi\pi$
self-energy diagram constrained to reproduce the observed width at the
physical pion mass. Including this contribution also provides a model
of the finite volume corrections arising from the infrared component
of the loop integral. In particular, we can also describe the lattice
results in the region $\mpi<m_\rho/2$, where the decay channel still
has higher energy because of momentum discretisation.
%

%
In this Letter
we present a global analysis of a very large set of sophisticated 
lattice simulation results. The aim is to produce an
accurate determination of the (real part of the) $\rho$-meson mass in
2-flavour QCD, with the required input of the experimental width. For a  
complete account of the analysis procedure, the reader is
referred to Ref.~\cite{Armour:2005}.

Partially-quenched lattice simulations are characterised by the
distinction between the masses of quarks coupling to external sources
and those associated with vacuum polarisation of the gauge
field. Subsequently, the construction of effective field theories
based on such simulations necessarily distinguishes the {\em valence-}
and {\em sea-} quark composition of hadronic states. While the
external legs of any $n$-point hadronic correlator are constructed of
valence quarks only, internal (hadron) loop diagrams may comprise
any mixture of sea and valence quarks. The mass of a
particular vector meson state is therefore described by
\begin{equation}
M_{ijk}^a = M(a,m_i;m_j,m_k)\, ,
\end{equation}
where the first two parameters of $M(a,m_i;m_j,m_k)$ specify the gauge
field ensemble with lattice spacing $a$ and sea-quark mass $m_{\rm
sea}=m_i$. The masses of the valence quarks are given by $m_j$ and
$m_k$. Similarly, $m_{ijk}^a$ is the mass of a pseudoscalar meson of
equivalent quark composition.

\begin{figure*}[t] 
\begin{center} 
\setlength{\unitlength}{1.0cm}
\setlength{\fboxsep}{0cm}
\begin{picture}(13.6,5.5)
\put( 0.00,2.00){\includegraphics[angle=0,width=4cm]{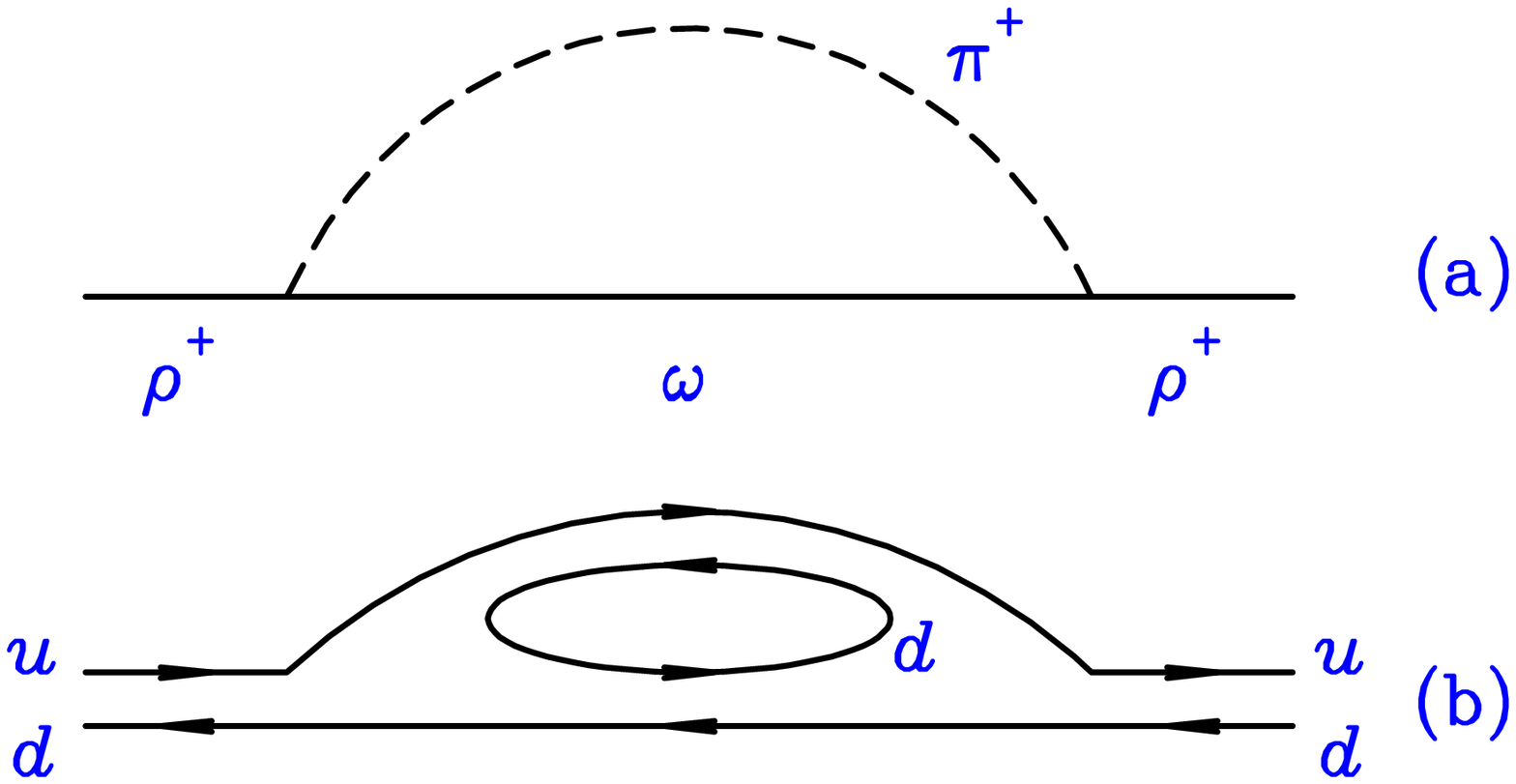}}
\put( 4.75,0.00){\includegraphics[angle=0,width=4cm]{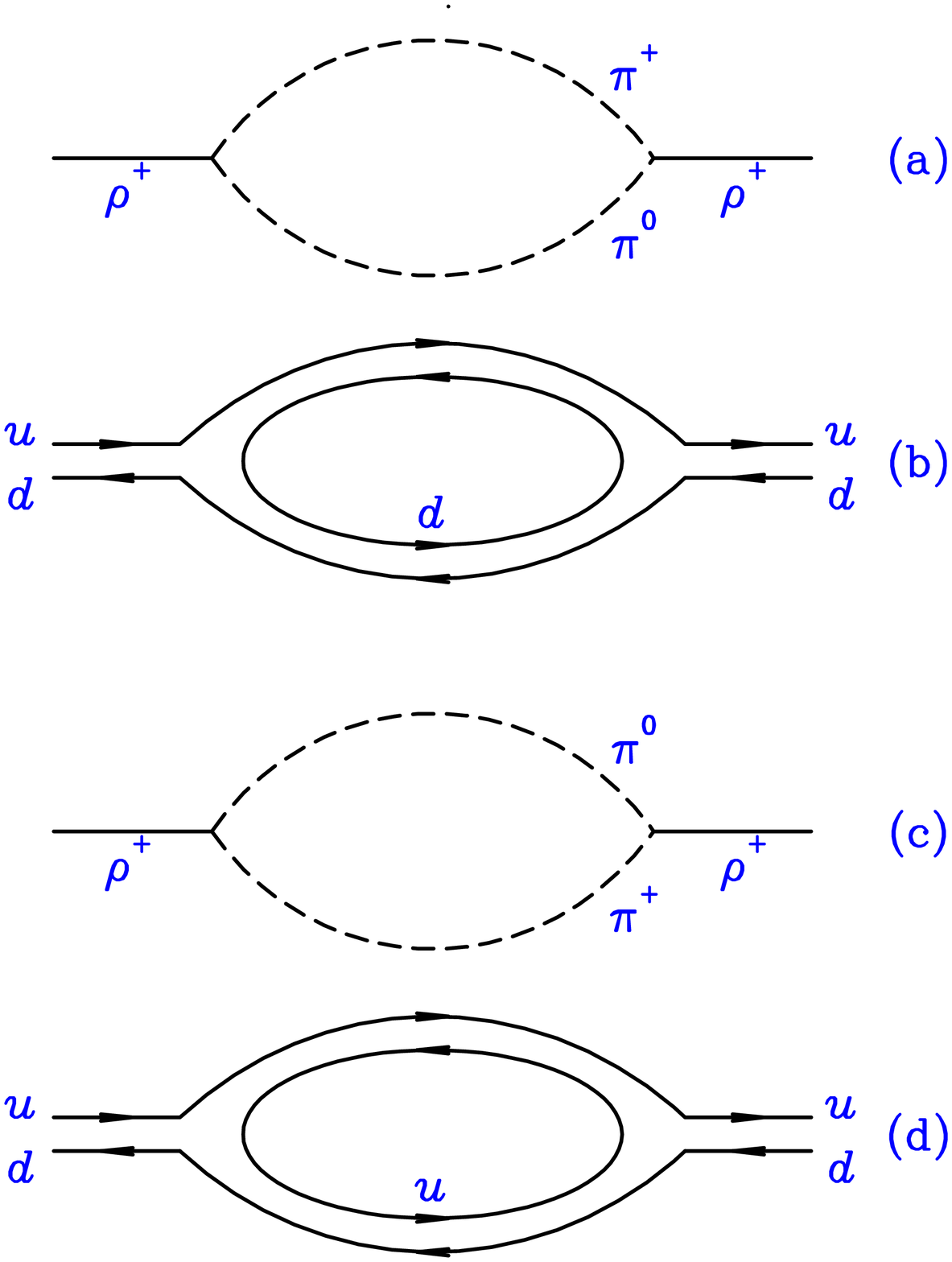}}
\put( 9.50,3.16){\includegraphics[angle=0,width=4cm]{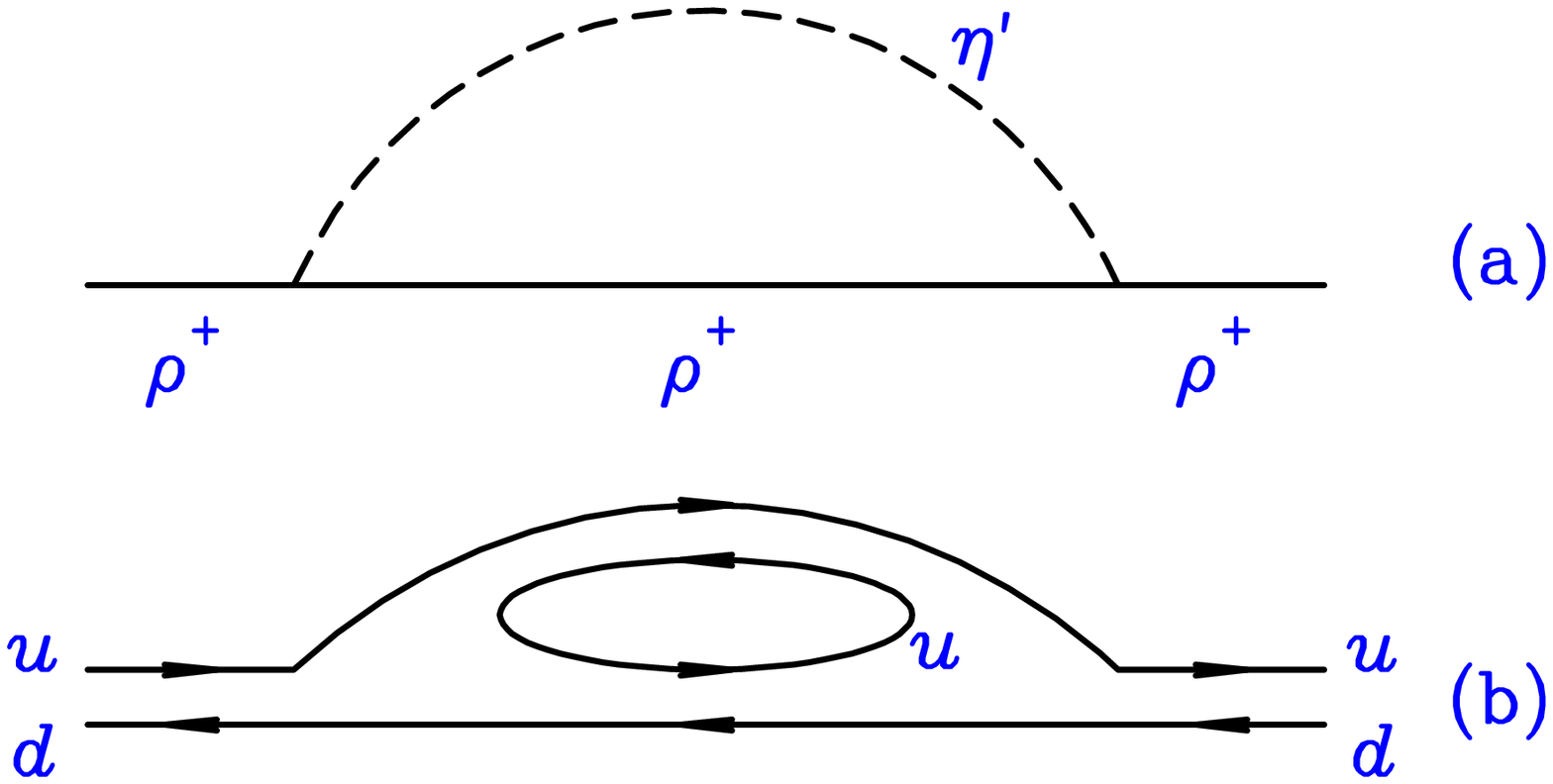}}
\put( 9.50,0.18){\includegraphics[angle=0,width=4cm]{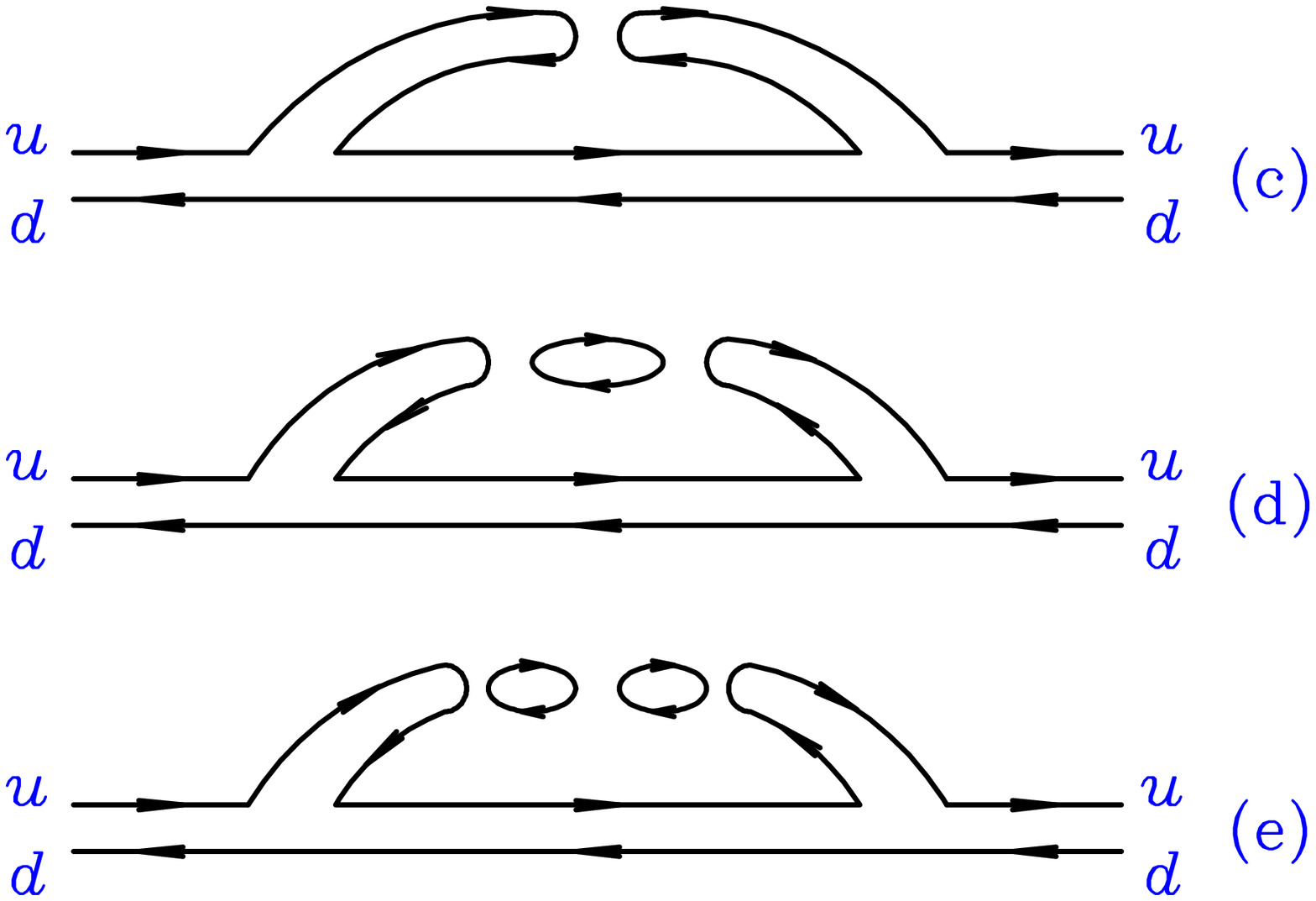}}
\end{picture}
\end{center} 
\caption{
Left: Diagram providing the leading nonanalytic contribution to the
  chiral expansion of the $\rho$-meson mass (a) and its associated
  quark-flow (b).
Middle: Two-pion contributions, (a), (c), to the $\rho$-meson
  self-energy and their associated quark-flow, (b), (d), respectively.
Right: The $\eta'$ contribution (a) and
  its associated quark flow diagrams in pQQCD.  While diagram (c)
  appears in quenched QCD, in pQQCD (or full QCD) it is complemented by an infinite series of
  terms, the first two of which are depicted in diagrams (d) and (e).
\label{quarkFlow}} 
\end{figure*}

The global parameterisation of the vector meson mass,
dependent on quark masses, lattice spacing and physical volume, is
written as
\begin{equation}
\label{eq:globalfit}
M_{ijj}^a{}^2 = (\alpha_0 + X_2 a^2 + \alpha_2 m_{ijj}^a{}^2 + \alpha_4 m_{ijj}^a{}^4)^2 + \Sigma^{\rm TOT}_{a;ijj}(L) \, .
\end{equation}
The total loop correction to the $M_{ijj}^a$ meson, on a finite box
of physical length $L=Na$, is described by
\begin{equation}
\Sigma^{\rm TOT}_{a;ijj}(L) = \Sigma_{\pi\pi}^\rho(m_{iij}^a,L) + \Sigma_{\pi\omega}^\rho(m_{iij}^a,L) 
                            + \Sigma_{\eta^\prime \rho}^\rho(m_{iij}^a,m_{ijj}^a,m_{iii}^a,L) \, .
\end{equation}
The relevant loop corrections are depicted in Fig.~\ref{quarkFlow}.  The
corresponding loop integrals, in the $L\to\infty$ limit, are given by
\begin{equation}
\label{eq:sigma_pipi}
\Sigma_{\pi\pi}^\rho(m_{iij}^a) = - \frac{f_{\rho\pi\pi}^2}{24\pi^{3}}
   \int \! d^3{k}~\frac{{k}^{2}u_{\pi\pi}^{2}(k)}
  {\sqrt{{k}^2+m_{iij}^a{}^2}\left[{k}^2+m_{iij}^a{}^2-\mu_\rho^2/4\right]} \, ,
\end{equation}
\begin{equation}
\label{eq:sigma_piomega}
\Sigma_{\pi\omega}^\rho(m_{iij}^a) = 
- \frac{f_{\rho \pi \omega}^2}{12\pi^{3} f_\pi^2}
   \int \! d^3{k}~\frac{{k}^{2}u^{2}(k)}
  {\sqrt{{k}^2+m_{iij}^a{}^2}\left[\sqrt{{k}^2+m_{iij}^a{}^2}+(M_{iij}^a-M_{ijj}^a)\right]} \, ,
\end{equation}
\begin{eqnarray}
\label{eq:sigma_dhp}
\Sigma^\rho_{\eta^\prime \rho}(m_{iij}^a,m_{ijj}^a,m_{iii}^a) &=&\nonumber\\
&&\hspace*{-35mm} 
\frac{f_{\rho \pi \omega}^2}{12\pi^{3} f_\pi^2}
   \int \! d^3{k}~\frac{{k}^{2}u^{2}(k)}{({k}^2+m_{ijj}^a{}^2)}
\bigg\{\frac{(m_{iij}^a{}^2 - m_{ijj}^a{}^2)}{({k}^{2} + m_{iij}^a{}^2)}
   +    \frac{(m_{iii}^a{}^2 - m_{ijj}^a{}^2)}{({k}^{2} + m_{ijj}^a{}^2)}\bigg\}\, .
\end{eqnarray}
Symbols are summarised as: $f_\pi=93\mev$; physical $\rho$ and $\pi$
masses, $\mu_\rho=0.770\gev$ and $\mu_\pi=0.140\gev$; $\rho\pi\pi$
coupling $f_{\rho\pi\pi}=6.028$ \cite{Leinweber:2001ac}; 
$f_{\rho \pi \omega}^2 = \mu_\rho\, g_2^2$;
$g_{2}$ 
(as introduced in Ref.~\cite{Chow:1997dw}) is related to the
$\omega\rho\pi$ coupling,
$g_{\omega\rho\pi}=16\gev^{-1}$ \cite{Lublinsky:1996yf}, by
$g_2=g_{\omega\rho\pi}\fpi/2=0.74$; $k=|\vec{k}|$; finite-range
regularisation is implemented with a dipole form,
$u(k)=(1+k^2/\Lambda^2)^{-2}$, and the $\rho\pi\pi$ coupling is preserved
at the physical threshold,
$u_{\pi\pi}(k)=u(k)u^{-1}(\sqrt{\mu_\rho^2/4-\mu_\pi^2})$.
%
%
By demanding this physical threshold the infrared behaviour is no
longer constrained and hence is not controlled by EFT. Therefore, such
a model is essential to extrapolate lattice results from the regime
$\mpi>m_\rho/2$.
%

The leading finite-volume corrections are trivially incorporated by
replacing the continuum loop integrals in
Eqs.~(\ref{eq:sigma_pipi},\ref{eq:sigma_piomega},\ref{eq:sigma_dhp})
by a sum over discrete momenta
\cite{Leinweber:2001ac,Young:2002cj,Young:2004tb}
\begin{equation}
\int\! d^3{k} \to \left(\frac{2\pi}{L}\right)^3 \sum_{\vec{k}}\, ,
\end{equation}
where $\vec{k}=(2\pi/L)\,\vec{i}$ for $\vec{i}\in\mathbb{Z}^3$. This
modification produces an infrared suppression of the loop integrals, and
is independent of the choice of ultra-violet regularisation
\cite{Beane:2004tw}.

The bracketed term in Eq.~(\ref{eq:globalfit}) describes the residual
variation of the vector meson mass which is not contained in the
one-loop Goldstone boson diagrams.  The analytic variation of the
quark-mass dependence is characterised by the continuum parameters,
$\alpha_i$. At finite lattice spacing, all terms at order $a$ and
$a^2$ must be treated consistently with the symmetry breaking patterns
of the prescribed fermion action \cite{Rupak:2002sm}. This can
potentially lead to more singular chiral behaviour in the effective
field theory at finite lattice spacing \cite{Aoki:2003yv}. In this
study, the leading lattice spacing corrections to the terms analytic
in the quark mass are investigated.


The lattice simulation data considered in this analysis come from
a large sample of partially quenched simulation results from the
CP-PACS Collaboration \cite{AliKhan:2001tx}. These simulations were 
performed using mean-field improved clover fermions at four different
couplings, $\beta$. For each value of the coupling, four different sea
quark masses have been calculated, yielding a total of 16 independent
gauge field ensembles. On each ensemble, the quark propagator has been
evaluated for five values of the valence quark mass. For the vector
mesons constructed of degenerate valence quarks there are a total of
80 ``data'' points. The lattice scale is set via the QCD Sommer scale
$r_0=0.49\fm$ \cite{sommer}, enabling all of these points to be shown in
physical units --- as in Fig.~\ref{fig:data}.

The form provided by Eq.~(\ref{eq:globalfit}) allows a universal fit
to all these points with just four free parameters (plus
regulator scale). This is therefore a highly constrained fit to the
large sample of simulation results. The best fit parameters are
displayed in Table~\ref{tab:fit}.
\begin{table}
\caption{The resulting global fit to the entire sample of lattice
results (with $\Lambda = 655\mev$).
\label{tab:fit}}
\begin{center}
\begin{tabular}{ccccc}
\hline\hline
$\alpha_0$        &      $X_2$        &    $\alpha_2$     &   $\alpha_4$       & $\chi^2/d.o.f.$ \\
(GeV)     &  (GeV-fm$^{-2}$)    &   (GeV$^{-1}$)    & (GeV$^{-3}$) &   \\
\hline
0.832\err{ 4}{ 4} & -1.40\err{ 3}{ 4} & 0.494\err{12}{11} & -0.061\err{ 8}{ 9} & 39 / 76\\
\hline\hline
\end{tabular}
\end{center}
\end{table}
The $\chi^2$ indicates that Eq.~(\ref{eq:globalfit}) accurately
describes this large quantity of data. The regulator mass, $\Lambda$,
has also been optimised to produce a best fit to the data, namely
$\Lambda=655 \pm 35\mev$, with the bound determined statistically from
the 1-$\sigma$ variation about the central fit. Studying the variation
of the fit over this domain introduces a small additional uncertainty
to the extrapolated result which is listed below in the error
estimate.

Variations of Eq.~(\ref{eq:globalfit}) have also been
investigated. Scaling corrections to the parameters $\alpha_2$ and
$\alpha_4$ yield coefficients which are consistent with zero.  Linear
corrections in the lattice spacing, taking the form of a term $X_1 a$,
are observed to be small and hence do not improve on the fit with
$X_1=0$.
Extending to higher analytic order in the quark mass expansion, by a
term $\alpha_6 \mpi^6$, reduces the stability of the fit, indicating
that the data are consistent with $\alpha_6=0$ --- see
Ref.~\cite{Armour:2005} for a complete account of these effects.
The systematic error quoted below covers the range found with 
all of these variations.

The fit parameters shown in Table~\ref{tab:fit} allow 
one to shift the simulation
results to the infinite-volume, continuum limit and
to remove the effects of partial quenching --- hence restoring unitarity in
the quark masses.  Complete details of the procedure are outlined in
Ref.~\cite{Armour:2005}. The results are displayed in
Fig.~\ref{fig:twist}, where we observe a remarkable result.
The tremendous spread of data seen in Fig.~\ref{fig:data} is 
dramatically reduced,
with all 80 points now lying very accurately on a universal curve.

The curve through Fig.~\ref{fig:twist} displays the determined
variation of the $\rho$-meson mass with pion mass.  This curve also
presents an extrapolation to the physical point, allowing extraction
of the physical $\rho$-meson mass
\begin{equation}
\label{eq:mass_final_adel}
M_\rho = 778(4)\er{16}{6}(8) \mev, \\
\end{equation}
where the first error is statistical, the second is systematic and the
third from the determination of $\Lambda$~\cite{Armour:2005}. This
result is in excellent agreement with the experimentally observed
mass.

The systematic uncertainty arises from the choice of fitting function,
as outlined above, and also from the choice of finite-range
regulator. In addition to the presented dipole form, the analysis has
been repeated with monopole and Gaussian regulators. The central
values of the extrapolated result with the monopole and Gaussian forms
differs by $+3$ and $-6\mev$, respectively, from the dipole
result. Each regularisation scheme produces a different model of the
$\rho\to\pi\pi$ vertex. This suggests that the model-dependence of
this contribution is small, once constrained to produce the correct
width at the physical point.

This extrapolation of the lattice results also offers an estimate of
the vector meson mass in the chiral limit, $M_\rho^0$. The central
value of the preferred fit gives a value $775\mev$, with errors
similar to those quoted above for the vector mass at the physical
point. These errors are from extrapolation of lattice results only,
whereas for phenomenological purposes the physical point provides a
further constraint. We therefore report the correlated difference
between the physical and chiral limit value
\begin{equation}
\label{eq:dM0}
M_\rho - M_\rho^0 = 3.7(2)\er{4}{4}(8) \mev\, . \\
\end{equation}
Interestingly, the chiral limit value of $M_\rho$ is very similar to
that of the physical value. This feature is observed in the reduction
in slope of the extrapolation curve in Figure \ref{fig:twist} as the
chiral limit is approached.  The underlying physics giving rise to
this reduced slope is the presence of substantial spectral strength in
the low-energy two-pion channel below the rho-meson mass
\cite{Leinweber:1993yw}.  This suggests a small sigma term for the
$\rho$ in comparison with the nucleon, where the curvature is enhanced
near the chiral limit \cite{Leinweber:2003dg}.


This analysis demonstrates the ability to treat all lattice artifacts
within a unified framework.  Both scaling violations and finite-volume
discrepancies can be removed through the procedure outlined.  The
number of simulation points can be increased dramatically by including
partially quenched results. This in turn permits a highly constrained
fit to produce an accurate extrapolation to the physical point.  With
minimal input, namely the $\rho\pi\pi$ and $\omega\rho\pi$ coupling
constants, the real part of the $\rho$-meson mass has been accurately
determined in two-flavour QCD\footnote{This is two-flavour QCD with
the $q\bar{q}$ force normalised to the physical value at a length
scale $r_0=0.49\fm$.}.  The final result for the pion mass variation,
as described by the universal curve in Fig.~\ref{fig:twist}, sets a
benchmark for the continuum, infinite-volume limit of the $\rho$ meson
in two-flavour QCD.

CRA and WA would like to thank the CSSM for their support and kind hospitality.
WA would like to thank PPARC for support.
The authors would like to thank W.~Melnitchouk, D.~Richards, G.~Shore and 
S.~Wright for helpful comments. 
This work was supported by the Australian Research Council
and by DOE contract DE-AC05-84ER40150, under which SURA operates
Jefferson Laboratory.


\begin{thebibliography}{99}

\bibitem{Aoki:2004ht}
  Y.~Aoki {\it et al.},
  arXiv:hep-lat/0411006.

\bibitem{Li:1971vr}
  L.~F.~Li and H.~Pagels,
  Phys.\ Rev.\ Lett.\  {\bf 26}, 1204 (1971).

\bibitem{Leinweber:1998ej}
  D.~B.~Leinweber, D.~H.~Lu and A.~W.~Thomas,
  Phys.\ Rev.\ D {\bf 60}, 034014 (1999)

\bibitem{Leinweber:1999ig}
  D.~B.~Leinweber, A.~W.~Thomas, K.~Tsushima and S.~V.~Wright,
  Phys.\ Rev.\ D {\bf 61}, 074502 (2000)

\bibitem{Borasoy:1996bx}
  B.~Borasoy and U.~G.~Meissner,
  Annals Phys.\  {\bf 254}, 192 (1997)

\bibitem{Young:2002ib}
  R.~D.~Young, D.~B.~Leinweber and A.~W.~Thomas,
  Prog.\ Part.\ Nucl.\ Phys.\  {\bf 50}, 399 (2003)

\bibitem{Leinweber:2003dg}
  D.~B.~Leinweber, A.~W.~Thomas and R.~D.~Young,
  Phys.\ Rev.\ Lett.\  {\bf 92} 242002 (2004)

\bibitem{AliKhan:2001tx}
  A.~Ali Khan {\it et al.}  [CP-PACS Collaboration],
  Phys.\ Rev.\ D {\bf 65}, 054505 (2002)
  [Erratum-ibid.\ D {\bf 67}, 059901 (2003)]

\bibitem{Gasser:1987zq}
  J.~Gasser and H.~Leutwyler,
  Nucl.\ Phys.\ B {\bf 307}, 763 (1988).

\bibitem{AliKhan:2003cu}
  A.~Ali Khan {\it et al.}  [QCDSF-UKQCD Collaboration],
  Nucl.\ Phys.\ B {\bf 689}, 175 (2004)

\bibitem{Procura:2003ig}
  M.~Procura, T.~R.~Hemmert and W.~Weise,
  Phys.\ Rev.\ D {\bf 69}, 034505 (2004)

\bibitem{Leinweber:2001ac}
D.~B.~Leinweber, A.~W.~Thomas, K.~Tsushima and S.~V.~Wright,
Phys.\ Rev.\ D {\bf 64}, 094502 (2001)

\bibitem{Young:2002cj}
  R.~D.~Young, D.~B.~Leinweber, A.~W.~Thomas and S.~V.~Wright,
  Phys.\ Rev.\ D {\bf 66}, 094507 (2002)

\bibitem{Young:2004tb}
  R.~D.~Young, D.~B.~Leinweber and A.~W.~Thomas,
  Phys.\ Rev.\ D {\bf 71}, 014001 (2005)

\bibitem{Colangelo:2003hf}
  G.~Colangelo and S.~Durr,
  Eur.\ Phys.\ J.\ C {\bf 33}, 543 (2004)

\bibitem{Beane:2004tw}
  S.~R.~Beane,
  Phys.\ Rev.\ D {\bf 70}, 034507 (2004)

\bibitem{Beane:2004rf}
  S.~R.~Beane and M.~J.~Savage,
  Phys.\ Rev.\ D {\bf 70}, 074029 (2004)

\bibitem{Leinweber:2004tc}
  D.~B.~Leinweber {\it et al.},
  Phys.\ Rev.\ Lett. {\bf 94}, 212001 (2005)

\bibitem{Borasoy:2004zf}
  B.~Borasoy and R.~Lewis,
  Phys.\ Rev.\ D {\bf 71}, 014033 (2005)

\bibitem{Luscher:1996ug}
  M.~Luscher, S.~Sint, R.~Sommer, P.~Weisz and U.~Wolff,
  Nucl.\ Phys.\ B {\bf 491}, 323 (1997)

\bibitem{Edwards:1997nh}
R.~G.~Edwards, U.~M.~Heller and T.~R.~Klassen,
  Phys.\ Rev.\ Lett.\  {\bf 80}, 3448 (1998)

\bibitem{Dong:2000mr}
  S.~J.~Dong, F.~X.~Lee, K.~F.~Liu and J.~B.~Zhang,
  Phys.\ Rev.\ Lett.\  {\bf 85}, 5051 (2000)

\bibitem{Zanotti:2004dr}
  J.~M.~Zanotti, B.~Lasscock, D.~B.~Leinweber and A.~G.~Williams,
  Phys.\ Rev.\ D {\bf 71}, 034510 (2005)

\bibitem{Rupak:2002sm}
  G.~Rupak and N.~Shoresh,
  Phys.\ Rev.\ D {\bf 66}, 054503 (2002)

\bibitem{Bar:2003mh}
  O.~B\"ar, G.~Rupak and N.~Shoresh,
  Phys.\ Rev.\ D {\bf 70}, 034508 (2004)

\bibitem{Aoki:2003yv}
  S.~Aoki,
  Phys.\ Rev.\ D {\bf 68}, 054508 (2003)

\bibitem{Beane:2003xv}
  S.~R.~Beane and M.~J.~Savage,
  Phys.\ Rev.\ D {\bf 68}, 114502 (2003)

\bibitem{Sharpe:2000bc}
  S.~R.~Sharpe and N.~Shoresh,
  Phys.\ Rev.\ D {\bf 62}, 094503 (2000)

\bibitem{Bijnens:1997ni}
  J.~Bijnens, P.~Gosdzinsky and P.~Talavera,
  Nucl.\ Phys.\ B {\bf 501}, 495 (1997)
  [arXiv:hep-ph/9704212].

\bibitem{Bruns:2004tj}
  P.~C.~Bruns and U.~G.~Meissner,
  Eur.\ Phys.\ J.\ C {\bf 40}, 97 (2005)
  [arXiv:hep-ph/0411223].

\bibitem{Armour:2005}
  W.~Armour {\it et al.}, in preparation.

\bibitem{Chow:1997dw}
  C.~K.~Chow and S.~J.~Rey,
  Nucl.\ Phys.\ B {\bf 528}, 303 (1998)

\bibitem{Lublinsky:1996yf}
  M.~Lublinsky,
  Phys.\ Rev.\ D {\bf 55}, 249 (1997)

\bibitem{sommer}
R.~Sommer,
Nucl.\ Phys.\ B {\bf 411}, 839 (1994).
R.~G.~Edwards, U.~M.~Heller and T.~R.~Klassen,
Nucl.\ Phys.\ B {\bf 517}, 377 (1998).

\bibitem{Leinweber:1993yw}
D.~B.~Leinweber and T.~D.~Cohen,
Phys.\ Rev.\ D {\bf 49} (1994) 3512
[arXiv:hep-ph/9307261].



\end{thebibliography}
\end{document}